\documentclass[10pt,conference]{IEEEtran}

\usepackage{url}

\usepackage{amsmath}
\usepackage{amssymb}

\newcommand{\abs}[1]{\left\lvert#1\right\rvert}

\newcommand{\R}{\mathbb{R}}
\newcommand{\X}{\{0,1\}^*}
\newcommand{\XI}{\{0,1\}^\infty}

\DeclareMathOperator{\Prob}{Pr}
\newcommand{\alp}{\mathcal{H}}

\newcommand{\noi}{\noindent}

\begin{document}

\title{A Statistical Mechanical Interpretation of Instantaneous Codes}

\author{
\authorblockN{Kohtaro Tadaki}
\authorblockA{Research and Development Initiative, Chuo University \\
1-13-27 Kasuga, Bunkyo-ku, Tokyo 112-8551, Japan \\
tadaki@kc.chuo-u.ac.jp}
}
%

\maketitle

\begin{abstract}
In this paper
we develop a statistical mechanical interpretation
of the noiseless source coding scheme based on
an absolutely optimal instantaneous code.
The notions in statistical mechanics such as
statistical mechanical entropy, temperature, and
thermal equilibrium
are translated into the context of noiseless source coding.
Especially,
it is discovered that
the temperature $1$ corresponds to the average codeword length of
an instantaneous code in this statistical mechanical interpretation of
noiseless source coding scheme.
This correspondence is also verified by the investigation
using box-counting dimension.
Using the notion of temperature and statistical mechanical arguments,
some information-theoretic relations can be derived in the manner
which appeals to intuition.
\end{abstract}

\section{Introduction}

We introduce a statistical mechanical interpretation
to the noiseless source coding scheme based on an absolutely optimal
instantaneous code.
The notions in statistical mechanics such as
statistical mechanical entropy, temperature, and
thermal equilibrium are translated
into the context of noiseless source coding.

We identify a coded message by an instantaneous code
with an energy eigenstate of a quantum system treated in statistical mechanics,
and the length of the coded message with the energy of the eigenstate.
The discreteness of the length of coded message
naturally corresponds to statistical mechanics
based on quantum mechanics and not on classical mechanics.
This is because the energy of a quantum system takes discrete value
while an energy takes continuous value in classical physics
in general.
Especially,
in this statistical mechanical interpretation of noiseless source coding,
the energy of the corresponding quantum system
is bounded to the above,
and therefore the system has negative temperature.
We discover that
the temperature $1$ corresponds to the average codeword length of
an instantaneous code
in the interpretation.
This correspondence is also verified by the investigation
based on box-counting dimension.

Note that,
we do not stick to the mathematical strictness of the argument
in this paper.
We respect the statistical mechanical intuition
in order to shed light on a hidden statistical mechanical aspect of
information theory,
and therefore make an argument
on the same level of mathematical strictness as statistical mechanics.

\section{Instantaneous codes}

We start with some notation on instantaneous codes
from information theory \cite{S48,A90,CT91}.

For any set $S$, $\#S$ denotes the number of elements in $S$.
We denote the set of all finite binary strings by $\X$.
For any $s\in\X$, $\abs{s}$ is the \textit{length} of $s$.
We define an \textit{alphabet} to be any nonempty finite set.

Let $X$ be an arbitrary random variable with an alphabet $\alp$
and a probability mass function $p_X(x)=\Prob\{X=x\}$, $x\in\alp$.
Then the \textit{entropy} $H(X)$ of $X$ is defined by
\begin{equation*}
  H(X)\equiv-\sum_{x\in\alp}p_X(x)\log p_X(x),
\end{equation*}
where the $\log$ is to the base $2$.
We will introduce the notion of a statistical mechanical entropy later.
Thus, in order to distinguish $H(X)$ from it,
we particularly call $H(X)$ the \textit{Shannon entropy} of $X$.
A subset $S$ of $\X$ is called a \textit{prefix-free set}
if no string in $S$ is a prefix of any other string in $S$.
An \textit{instantaneous code} $C$ for the random variable $X$
is an injective mapping from $\alp$ to $\X$
such that $C(\alp)\equiv\{C(x)|x\in\alp\}$ is a prefix-free set.
For each $x\in\alp$,
$C(x)$ is called the \textit{codeword} corresponding to $x$
and $\abs{C(x)}$ is denoted by $l(x)$.
A sequence $x_1,x_2,\dots,x_N$ with $x_i\in\alp$ is called a $message$.
On the other hand,
the finite binary string $C(x_1)C(x_2)\dotsm C(x_N)$ is called
the \textit{coded message} for a message $x_1,x_2,\dots,x_N$.

An instantaneous code play an important role in
the noiseless source coding problem described as follows.
Let $X_1,X_2,\dots,X_N$ be independent identically distributed
random variables drawn from the probability mass function $p_X(x)$.
The objective of the noiseless source coding problem is
to minimize the length of the binary string $C(x_1)C(x_2)\dotsm C(x_N)$
for a message $x_1,x_2,\dots,x_N$ generated by the random variables $\{X_i\}$
as $N\to\infty$.
For that purpose,
it is sufficient to consider
the \textit{average codeword length} $L_X(C)$ of an instantaneous code $C$
for the random variable $X$,
which is defined by
\begin{equation*}
  L_X(C)\equiv\sum_{x\in\alp}p_X(x)l(x)
\end{equation*}
independently on the value of $N$.
We can then show that
$L_X(C)\ge H(X)$
for any instantaneous code $C$ for the random variable $X$.
Hence, the Shannon entropy gives the data compression limit
for the noiseless source coding problem based on instantaneous codes.
Thus, it is important to consider the notion of
absolutely optimality of an instantaneous code,
where we say that
an instantaneous code $C$ for the random variable $X$ is
\textit{absolutely optimal} if $L_X(C)=H(X)$.
We can see that
an instantaneous code $C$ is absolutely optimal
if and only if $p_X(x)=2^{-l(x)}$ for all $x\in\alp$.

Finally, for each $x^N=(x_1,x_2,\dots,x_N)\in\alp^N$,
we define $p_X(x^N)$ as $p_X(x_1)p_X(x_2)\dotsm p_X(x_N)$.

\section{Statistical Mechanical Interpretation}

In this section,
we develop a statistical mechanical interpretation
of the noiseless source coding by an instantaneous code.
In what follows,
we assume that an instantaneous code $C$ for a random variable $X$
is absolutely optimal.

In statistical mechanics \cite{Re65,TKS92,Ru99},
we consider a quantum system $\mathcal{S}_{\text{total}}$
which consists in a large number of identical quantum subsystems.
Let $N$ be a number of such subsystems.
For example, $N\sim 10^{22}$
for
$1\,\mathrm{cm^3}$
of a gas at room temperature.
We assume here that each quantum subsystem can be distinguishable from others.
Thus, we deal with quantum particles which obey
Maxwell-Boltzmann statistics and not
Bose-Einstein statistics or Fermi-Dirac statistics.
Under this assumption,
we can identify the $i$th quantum subsystem $\mathcal{S}_i$
for each $i=1,\dots,N$.
In quantum mechanics,
any quantum system is described by a quantum state completely.
In statistical mechanics,
among all quantum states,
energy eigenstates are of particular importance.
Any energy eigenstate of each subsystem $\mathcal{S}_i$ can be specified by
a number $n=1,2,3,\dotsc$, called a \textit{quantum number},
where the subsystem in the energy eigenstate specified by $n$ has
the energy $E_n$.
Then, any energy eigenstate of the system $\mathcal{S}_{\text{total}}$
can be specified by an $N$-tuple $(n_1,n_2,\dots,n_N)$ of
quantum numbers.
If the state of the system $\mathcal{S}_{\text{total}}$ is
the energy eigenstate specified by $(n_1,n_2,\dots,n_N)$,
then the state of each subsystem $\mathcal{S}_i$ is
the energy eigenstate specified by $n_i$
and the system $\mathcal{S}_{\text{total}}$
has the energy $E_{n_1}+E_{n_2}+ \dots +E_{n_N}$.
Then,
the fundamental postulate of statistical mechanics
is stated as follows.
\medskip

\noi
\textbf{Fundamental Postulate:}
If the energy of the system $\mathcal{S}_{\text{total}}$ is known to have
a constant value in the range between $E$ and $E+\delta E$,
where $\delta E$ is the indeterminacy
in measurement of the energy of the system $\mathcal{S}_{\text{total}}$,
then the system $\mathcal{S}_{\text{total}}$ is equally likely to be in
any energy eigenstate specified by
$(n_1,n_2,\dots,n_N)$
such that $E\le E_{n_1}+E_{n_2}+ \dots +E_{n_N}\le E+\delta E$.
\medskip

Let $\Omega(E,N)$ be the total number of energy eigenstates
of $\mathcal{S}_{\text{total}}$ specified by $(n_1,n_2,\dots,n_N)$
such that $E\le E_{n_1}+E_{n_2}+ \dots +E_{n_N}\le E+\delta E$.
The above postulate states that
any energy eigenstate of $\mathcal{S}_{\text{total}}$
whose energy lies between $E$ and $E+\delta E$
occurs with the probability $1/\Omega(E,N)$.
This uniform distribution of energy eigenstates
whose energy lies between $E$ and $E+\delta E$ is called
a \textit{microcanonical ensemble}.
In statistical mechanics, the \textit{entropy} $S(E,N)$ of
the system $\mathcal{S}_{\text{total}}$ is then defined by
\begin{equation*}
  S(E,N)\equiv k\ln \Omega(E,N),
\end{equation*}
where $k$ is a positive constant, called the \textit{Boltzmann Constant},
and the $\ln$ denotes
the natural logarithm.
Note that,
in statistical mechanics,
the entropy $S(E,N)$ is normally estimated to first order in $N$ and $E$.
Thus the magnitude of the indeterminacy $\delta E$ of the energy does
not matter unless it is too small.
The \textit{temperature} $T(E,N)$ of the system $\mathcal{S}_{\text{total}}$
is defined by
\begin{equation*}
  \frac{1}{T(E,N)}\equiv \frac{\partial S}{\partial E}(E,N).
\end{equation*}
Thus the temperature is a function of $E$ and $N$.
The average energy $\varepsilon$ per one subsystem is given by $E/N$.

Now we give a statistical mechanical interpretation
to the noiseless source coding scheme based on an instantaneous code.
Let $X$ be an arbitrary random variable with an alphabet $\alp$,
and let $C$ be an absolutely optimal instantaneous code
for the random variable $X$.
Let $X_1,X_2,\dots,X_N$ be independent identically distributed
random variables drawn from the probability mass function $p_X(x)$
for a large $N$, say $N\sim 10^{22}$.
We relate the noiseless source coding based on $C$ to
the above statistical mechanics as follows.
The sequence $X_1,X_2,\dots,X_N$ corresponds to
the quantum system $\mathcal{S}_{\text{total}}$,
where each $X_i$ corresponds to the $i$th quantum subsystem $\mathcal{S}_i$.
We relate $x\in\alp$, or equivalently, $C(x)$
to an energy eigenstate of a subsystem,
and we relate $l(x)=\abs{C(x)}$ to an energy $E_n$ of
the energy eigenstate of the subsystem.
Then a sequence $(x_1,\dots,x_N)\in\alp^N$,
or equivalently, a finite binary string $C(x_1)\dotsm C(x_N)$
corresponds to an energy eigenstate of $\mathcal{S}_{\text{total}}$
specified by $(n_1,\dots,n_N)$.
Thus, $l(x_1)+\dots+l(x_N)=\abs{C(x_1)\dotsm C(x_N)}$ corresponds to
the energy $E_{n_1}+ \dots +E_{n_N}$ of
the energy eigenstate of $\mathcal{S}_{\text{total}}$.

We define a subset $C(L,N)$ of $\X$ as the set of all coded messages
$C(x_1)\dotsm C(x_N)$ whose length lies between $L$ and $L+\delta L$.
Then $\Omega(L,N)$ is defined as $\#C(L,N)$.
Therefore $\Omega(L,N)$ is the total number of coded messages
whose length lies between $L$ and $L+\delta L$.
We can see that
if $C(x_1)\dotsm C(x_N)\in C(L,N)$,
then $2^{-L}\le p(x^N)\le 2^{-(L+\delta L)}$.
This is because $C$ is an absolutely optimal instantaneous code.
Thus all coded messages $C(x_1)\dotsm C(x_N)\in C(L,N)$ occur
with the probability $2^{-L}$.
Note here that we care nothing about the magnitude of $\delta L$,
as in the case of statistical mechanics.
Thus,
given that the length of coded message is $L$,
all coded messages occur with the same probability $1/\Omega(L,N)$.
We introduce a micro-canonical ensemble
on the noiseless source coding in this manner.
Thus we can develop a certain sort of statistical mechanics
on the noiseless source coding scheme.

The \textit{statistical mechanical entropy} $S(L,N)$ of
the instantaneous code $C$ is defined by
\begin{equation*}
  S(L,N)\equiv \log \Omega(L,N).
\end{equation*}
The \textit{temperature} $T(L,N)$ of $C$ is then defined by
\begin{equation*}
  \frac{1}{T(L,N)}\equiv \frac{\partial S}{\partial L}(L,N).
\end{equation*}
Thus the temperature is a function of $L$ and $N$.
The average length $\lambda$ of coded message per one codeword
is given by $L/N$.
The average length $\lambda$ corresponds to the average energy $\varepsilon$
in the statistical mechanics above.

\section{Properties of Statistical Mechanical Entropy}
\label{psme}

In statistical mechanics,
it is important to know the values of the energy $E_n$ of
subsystem $\mathcal{S}_i$ for all quantum numbers $n$,
since the values determine the entropy $S(E,N)$ of the quantum system
$\mathcal{S}_{\text{total}}$.
Corresponding to this fact,
the knowledge of $l(x)$ for all $x\in\alp$ is important to calculate $S(L,N)$.
We investigate some properties of $S(L,N)$ and $T(L,N)$ based on
$l(x)$ in the following.

As is well known in statistical mechanics,
if the energy of a quantum system $\mathcal{S}_{\text{total}}$ is
bounded to the above, then the system can have negative temperature.
The same situation happens in our statistical mechanics developed on
an instantaneous code $C$,
since there are only finite codewords of $C$.
We define $l_{\text{min}}$ and $l_{\text{max}}$ as
$\min\{l(x)\mid x\in\alp\}$ and $\max\{l(x)\mid x\in\alp\}$, respectively.
Given $N$,
the statistical mechanical entropy $S(L,N)$ is a unimodal function of $L$
and takes nonzero value only between $Nl_{\text{min}}$ and $Nl_{\text{max}}$.
Let $L_0$ be the value $L$ which maximizes $S(L,N)$.
If $L<L_0$ then $T(L,N)>0$.
On the other hand, if $L>L_0$ then $T(L,N)<0$.
The temperature $T(L,N)$ takes $\pm\infty$ at $L=L_0$.

According to the method of Boltzmann and Planck (see e.g.~\cite{TKS92}),
we can show that
\begin{equation}\label{BP}
  S(L,N)=NH(G(C,T(L,N))),
\end{equation}
where $G(C,T)$ is the random variable with the alphabet $\alp$
and the probability mass function $p_{G(C,T)}(x)=\Prob\{G(C,T)=x\}$
defined by
\begin{equation*}
  p_{G(C,T)}(x)\equiv
  \frac{2^{-l(x)/T}}{\sum_{a\in\alp}2^{-l(a)/T}}.
\end{equation*}
The temperature $T(L,N)$ is implicitly determined
through
the equation
\begin{equation}\label{al}
  \frac{L}{N}=\sum_{x\in\alp} l(x)p_{G(C,T(L,N))}(x)
\end{equation}
as a function of $L$ and $N$.
These properties of $S(L,N)$ and $T(L,N)$ are derived
only based on a combinatorial aspect of $S(L,N)$.

Now, let us take into account the probabilistic issue
given by the random variables $X_1,X_2,\dots,X_N$.
Since the instantaneous code $C$ is absolutely optimal,
a particular coded message of length $L$ occurs with probability $2^{-L}$.
Thus the probability that some coded message of length $L$ occurs
is given by $2^{-L}\Omega(L,N)$.
Hence,
by differentiating $2^{-L}\Omega(L,N)$ on $L$ and setting the result to $0$,
we can determine the most probable length $L^*$ of coded message, given $N$.
Thus we have the relation
\begin{equation*}
  \frac{\partial}{\partial L}\{-L+ S(L,N)\}\Big|_{(L,N)=(L^*,N)}=0,
\end{equation*}
which is satisfied by $L^*$.
It follows that $T(L^*,N)=1$,
Thus, the temperature $1$ corresponds to the most probable length $L^*$.
On the other hand,
$p_{G(C,1)}(x)=2^{-l(x)}$ at $T(L^*,N)=1$,
and therefore, by \eqref{al}, we have $L^*/N=H(X)=L_X(C)$.
Since $C$ is absolutely optimal,
this result is consistent with the law of large numbers.
Thus, the temperature $1$ corresponds to the average codeword length $L_X(C)$,
which is equal to the average length $\lambda$ of coded message
per one codeword at the temperature $1$.

\section{Thermal Equilibrium between Two Instantaneous Codes}
\label{te}

Let $X^{\text{I}}$ be an arbitrary random variable with an alphabet
$\alp^{\text{I}}$,
and let $C^{\text{I}}$ be an absolutely optimal instantaneous code
for the random variable $X^{\text{I}}$.
Let $X^{\text{I}}_1,X^{\text{I}}_2,\dots,X^{\text{I}}_{N^{\text{I}}}$ be
independent identically distributed random variables drawn
from the probability mass function $p_{X^{\text{I}}}(x)$
for a large $N^{\text{I}}$.
On the other hand,
let $X^{\text{II}}$ be an arbitrary random variable with an alphabet
$\alp^{\text{II}}$,
and let $C^{\text{II}}$ be an absolutely optimal instantaneous code
for the random variable $X^{\text{II}}$.
Let $X^{\text{II}}_1,X^{\text{II}}_2,\dots,X^{\text{II}}_{N^{\text{II}}}$ be
independent identically distributed random variables drawn
from the probability mass function $p_{X^{\text{II}}}(x)$
for a large $N^{\text{II}}$.

Consider the following problem:
Find the most probable values $L^{\text{I}}$ and $L^{\text{II}}$,
given that the sum $L^{\text{I}}+L^{\text{II}}$ of 
the length $L^{\text{I}}$ of coded message by $C^{\text{I}}$
for the random variables $\{X^{\text{I}}_i\}$
and
the length $L^{\text{II}}$ of coded message by $C^{\text{II}}$
for the random variables $\{X^{\text{II}}_j\}$
is equal to $L$.

In order to solve this problem,
the statistical mechanical notion of ``thermal equilibrium'' can be used.
We first note that
a particular coded message by $C^{\text{I}}$ of length $L_{\text{I}}$ and
a particular coded message by $C^{\text{II}}$ of length $L_{\text{II}}$
occur with probability $2^{-L_{\text{I}}}2^{-L_{\text{II}}}=2^{-L}$,
since the instantaneous codes $C^{\text{I}}$ and $C^{\text{II}}$ are
absolutely optimal.
Thus,
any particular pair of coded messages by $C^{\text{I}}$ and $C^{\text{II}}$
occurs with an equal probability,
given that the total length of coded messages
for $\{X^{\text{I}}_i\}$ and $\{X^{\text{II}}_j\}$ is $L$.
Therefore,
the most probable allocation
${L_{\text{I}}}^*$ and ${L_{\text{II}}}^*$ of $L=L_{\text{I}}+L_{\text{II}}$
maximizes the product
$\Omega_{\text{I}}(L_{\text{I}},N_{\text{I}})
\Omega_{\text{II}}(L_{\text{II}},N_{\text{II}})$.
We see that this condition is equivalent to the equality:
\begin{equation*}
  T_{\text{I}}(L_{\text{I}}^*,N_{\text{I}})=
  T_{\text{II}}(L_{\text{II}}^*,N_{\text{II}}),
\end{equation*}
where the functions $T_{\text{I}}$ and $T_{\text{II}}$ are the temperature
of $C^{\text{I}}$ and $C^{\text{II}}$, respectively.
This equality corresponds to the condition on the thermal equilibrium between
two systems, given a total energy, in statistical mechanics.
Using \eqref{al},
the value of $T_{\text{I}}(L_{\text{I}}^*,N_{\text{I}})=
T_{\text{II}}(L_{\text{II}}^*,N_{\text{II}})$ is obtained by solving
the equation on $T$:
\begin{eqnarray*}
  &&\frac{N^{\text{I}}}{L}
  \sum_{x\in\alp^{\text{I}}}
  \abs{C^{\text{I}}(x)}p_{G(C^{\text{I}},T)}(x)+\\
  &&\frac{N^{\text{II}}}{L}
  \sum_{x\in\alp^{\text{II}}}
  \abs{C^{\text{II}}(x)}p_{G(C^{\text{II}},T)}(x)\\
  &&=1.
\end{eqnarray*}
Then, again by \eqref{al},
the most probable values $L_{\text{I}}^*$ and $L_{\text{II}}^*$
are determined.

\section{Dimension of Coded Messages}

The notion of dimension plays an important role in fractal geometry \cite{F90}.
In this section,
we investigate our statistical mechanical interpretation of
the noiseless source coding from the point of view of dimension.
Let $F$ be a bounded subset of $\R$, and let $N_n(F)$ be
the number of $2^{-n}$-mesh cubes that intersect $F$,
where $2^{-n}$-mesh cube is a subset of $\R$
in the form of $[m2^{-n},(m+1)2^{-n}]$ for some integer $m$.
The \textit{box-counting dimension} $\dim_B F$ of $F$ is then defined by
\begin{equation*}
  \dim_B F \equiv \lim_{n\to\infty} \frac{\log N_n(F)}{n}.
\end{equation*}
Let $\XI=\{b_1b_2b_3\dotsm\mid b_i=0,1\text{ for all }i=1,2,3,\dotsc\}$ be
the set of all infinite binary strings.
In \cite{T02}
we investigate the dimension of sets of
coded messages of infinite length,
where the number of distinct codewords is finite or infinite.
In a similar manner,
we investigate the set of coded messages of infinite length by
an absolutely optimal instantaneous code $C$.

By \eqref{al},
the ratio $L/N$ is uniquely determined by temperature $T$.
Thus,
by letting $L,N\to\infty$ while keeping the ratio $L/N$ constant,
we can regard the set $C(L,N)$ as a subset of $\XI$.
This kind of limit is called
the \textit{thermodynamic limit} in statistical mechanics.
Taking the thermodynamic limit,
we denote $C(L,N)$ by $F(T)$,
where $T$ is related to the limit value of $L/N$ through \eqref{al}.
Although $F(T)$ is a subset of $\XI$,
we can regard $F(T)$ as a subset of $[0,1]$ by identifying $\alpha\in\XI$ with
the real number $0.\alpha$.
In this manner,
we can consider the box-counting dimension $\dim_B F(T)$ of $F(T)$.

We investigate the dependency of $\dim_B F(T)$ on temperature $T$ with
$-\infty\le T\le \infty$.
First it can be shown that
\begin{eqnarray*}
  \dim_B F(T)
  &=&\lim_{L,N\to\infty} \frac{\log \Omega(L,N)}{L}\\
  &=&\lim_{L,N\to\infty} \frac{S(L,N)}{L},
\end{eqnarray*}
where the limits are taken while satisfying \eqref{al} for each $T$.
Thus the statistical mechanical entropy $S(L,N)$ and
the box-counting dimension $\dim_B F(T)$ of $F(T)$ are closely related.
By \eqref{BP} and \eqref{al},
we can
obtain,
as an explicit formula of $T$,
\begin{equation}\label{fd}
  \dim_B F(T)=
  \frac{1}{T}+\frac{1}{\lambda(T)}\log \sum_{x\in\alp} 2^{-l(x)/T},
\end{equation}
where $\lambda(T)$ is defined by
\begin{equation*}
  \lambda(T)\equiv\sum_{x\in\alp} l(x)p_{G(C,T)}(x).
\end{equation*}
We define the ``degeneracy factors'' $d_{\text{min}}$ and $d_{\text{max}}$
of the lowest and highest ``energies'' by
$d_{\text{min}}\equiv\#\{x\in\alp\mid\l(x)=l_{\text{min}}\}$ and
$d_{\text{max}}\equiv\#\{x\in\alp\mid\l(x)=l_{\text{max}}\}$, respectively.
Note here that since $C$ is assumed to be absolutely optimal,
$\sum_{x\in\alp}2^{-l(x)}=1$
and therefore $d_{\text{max}}$ can be shown to be an even number.
In the increasing order of the ratio $L/N$ (i.e. $\lambda(T)$),
we see from \eqref{fd} that
\begin{eqnarray*}
  \lim_{T\to +0} \dim_B F(T)&=&\frac{\log d_{\text{min}}}{l_{\text{min}}},\\
  \dim_B F(1)&=&1,\\
  \lim_{T\to \pm\infty} \dim_B F(T)&=&\frac{n\log n}{\sum_{x\in\alp}l(x)},\\
  \lim_{T\to -0} \dim_B F(T)&=&\frac{\log d_{\text{max}}}{l_{\text{max}}}.
\end{eqnarray*}
We can show that $n\log n<\sum_{x\in\alp}l(x)$
unless all codewords have the same length,
and obviously $\log d_{\text{min}}/l_{\text{min}}<1$ and
$\log d_{\text{max}}/l_{\text{max}}<1$ except for such a trivial case.
Thus, in general,
the dimension $\dim_B F(T)$ is maximized at the temperature $T=1$.
This can be checked using \eqref{fd} based on
the differentiation of $\dim_B F(T)$.
That is,
we can show that,
if all codewords do not have the same length,
then the following hold:
\vspace*{2mm}
\begin{enumerate}
 \item $\displaystyle\frac{d}{dT}\dim_B F(T)\Big|_{T=T_0}=0$
   if and only if $T_0=1$,
 \vspace*{3mm}
 \item $\displaystyle\frac{d^2}{dT^2}\dim_B F(T)\Big|_{T=1}<0$.
 \vspace*{2mm}
\end{enumerate}
Note that all coded messages $C(x_1)C(x_2)\dotsm$
of infinite length form the set $\XI$ and therefore the interval $[0,1]$,
since $C$ is an absolutely optimal instantaneous code.
Thus, since $\dim_B F(1)$ is equal to $\dim_B [0,1]$,
the set $F(1)$ is as rich as the set $[0,1]$ in a certain sense.
This can be explained as follows.
Since $L/N=L_X(C)$ at the temperature $T=1$, as seen in Section \ref{psme},
by the law of large numbers,
the length of coded message for a message of length $N$ is
likely to equal $NL_X(C)$, for a sufficiently large $N$.
Thus $F(1)$ contains almost all finite binary string of length $L$.
In other words,
$F(1)$ consists in coded messages for all messages which
form the typical set in a sense.

\section{Conclusion}

In this paper
we have developed a statistical mechanical interpretation
of the noiseless source coding scheme based on
an absolutely optimal instantaneous code.
The notions in statistical mechanics such as
statistical mechanical entropy, temperature, and
thermal equilibrium are translated into the context of information theory.
Especially,
it is discovered that
the temperature $1$ corresponds to the average codeword length $L_X(C)$
in this statistical mechanical interpretation of information theory.
This correspondence is also verified by the investigation
using box-counting dimension.
The argument is not necessarily mathematically rigorous.
However,
using the notion of temperature and statistical mechanical arguments,
several information-theoretic relations can be derived in the manner
which appeals to intuition.

A statistical mechanical interpretation of
the general case where
the underlying instantaneous code is not necessarily absolutely optimal
is reported in another work.


\section*{Acknowledgments}

The author is grateful to
the 21st Century COE Program and the Research and Development Initiative
of Chuo University for the financial supports.




%

\end{document}